\documentclass[preprint,12pt, a4paper]{elsarticle}

\usepackage{amssymb}
\usepackage{xcolor}

\usepackage{lineno}

\usepackage{listings}
\usepackage{multirow}
\usepackage{colortbl}
\usepackage{makecell}

\usepackage{float}
\restylefloat{table}
\usepackage{caption}
\usepackage{subcaption}

\newcommand{\CC}{C\nolinebreak\hspace{-.05em}\raisebox{.4ex}{\tiny\bf +}\nolinebreak\hspace{-.10em}\raisebox{.4ex}{\tiny\bf +}}

\journal{SoftwareX}
\usepackage{hyperref}

\begin{document}

\begin{frontmatter}



\title{PRETUS: A plug-in based platform for real-time ultrasound imaging research}

\author[label1]{Alberto Gomez}
\author[label1,label2]{Veronika A. Zimmer}
\author[label1]{Gavin Wheeler}
\author[label1]{Nicolas Toussaint}
\author[label1]{Shujie Deng}
\author[label1]{Robert Wright}
\author[label1]{Emily Skelton}
\author[label1]{Jackie Matthew}
\author[label3,label5]{Bernhard Kainz}
\author[label1]{Jo Hajnal}
\author[label1,label2,label4]{Julia Schnabel}

\address[label1]{School of Biomedical Engineering \& Imaging Sciences, King's College London, UK}
\address[label2]{Department of Informatics, Technical University Munich, Germany}
\address[label3]{Department of Computing, Imperial College London, UK}
\address[label5]{Friedrich-Alexander-University Erlangen-N\"urnberg, Germany}
\address[label4]{Helmholtz Zentrum München – German Research Center for Environmental Health, Germany}

\begin{abstract}

We present PRETUS -- a Plugin-based Real Time UltraSound  software platform for live ultrasound image analysis and operator support. The software is lightweight; functionality is brought in via independent plug-ins that can be arranged in sequence. The software allows to capture the real-time stream of ultrasound images from virtually any ultrasound machine, applies computational methods and visualises the results on-the-fly. 

Plug-ins can run concurrently without blocking each other. They can be implemented in \CC and Python. A graphical user interface can be implemented for each plug-in, and presented to the user in a compact way. The software is free and open source, and allows for rapid prototyping and testing of real-time ultrasound imaging methods in a manufacturer-agnostic fashion. The software is provided with input, output and processing plug-ins, as well as with tutorials to illustrate how to develop new plug-ins for PRETUS.
\end{abstract}


\end{frontmatter}

\begin{table}[H]
\begin{tabular}{|l|p{6.5cm}|p{6.5cm}|}
\hline
\textbf{Nr.} & \textbf{Code metadata description} & \textbf{Please fill in this column} \\
\hline
C1 & Current code version & 1.0 \\
\hline
C2 & Permanent link to code/repository used for this code version & $https://github.com/gomezalberto/pretus$ \\
\hline
C4 & Legal Code License   & MIT License \\
\hline
C5 & Code versioning system used & git \\
\hline
C6 & Software code languages, tools, and services used & \CC, Python \\
\hline
C7 & Compilation requirements, operating environments \& dependencies & Qt, ITK, VTK, Boost, OpenCV (specific plug-ins may have additional dependencies) \\
\hline
C8 & If available Link to developer documentation/manual & N/A (can be Doxygen generated) \\
\hline
C9 & Support email for questions & pretus@googlegroups.com\\
\hline
\end{tabular}
\caption{Code metadata (mandatory)}
\label{} 
\end{table}








\section{Motivation and significance}
\label{sec:motivation-significance}

Ultrasound (US) imaging is one of the most widely used medical imaging modalities, because it is portable, affordable and safe, and can be used to gain insight about most body organs. Moreover, US is, unlike other common modalities such as Computed Tomography (CT), Magnetic Resonance Imaging (MRI) or X-ray, a real-time modality by design: to use an ultrasound system the operator needs to interpret the real-time stream of images shown on the display and use the interpreted information to guide the US transducer to the desired view. As the examination progresses, the operator typically stores a few tens of static images or short clips for reporting or further investigation. Importantly, the main clinical use of US images is during the procedure. This is because in diagnostic imaging, diagnosis is done by the operator as the images are being acquired and interpreted. In interventional imaging, surgical tools are guided using images in real-time image.

US image analysis is a very active area of research~\cite{Che2017,Meiburger2018,Liu2019}, and most published work has focused in the `offline' analysis of images and clips stored by the operator as described above. However, real-time analysis of US image streams can potentially transform the way ultrasound is utilised since it can provide the operator with extended information and guidance \emph{during} the examination, which as pointed out before offers the biggest potential benefit.

We identified three main reasons why limited work has been done on real-time US image analysis: firstly, collecting real-time data is not supported by most US systems and requires external equipment, such as a video framegrabber; the few systems which do support real-time streaming of DICOM (Digital Imaging and Communications in Medicine, an international standard for storage and transmission of medical images - \url{https://www.dicomstandard.org}) frames require a proprietary protocol to access the image stream. Secondly, existing research tools used for implementing real-time image analysis methods are designed to perform a single computational task on the stream of images, however real-time analysis often requires a number of tasks to run in succession (or in parallel) without compromising the real-time performance. And thirdly, in order to carry out translational research, the results of the real-time analysis must be shown to the operator in a way that does not require switching between displays during the scan, as this would be unfeasible. In this paper we describe \emph{PRETUS: Plugin-based Real Time UltraSound}, a software that addresses these three challenges while remaining a simple, lightweight tool that can be easily extended via plug-ins-- independent pieces of software that can be built separately to the main software and can be added dynamically to extend its functionality.

A number of research softwares have been proposed over the last years supporting real-time ultrasound imaging for research purposes. Of those, the most widely used are Slicer IGT \cite{ungi2016open} and MITK IGT \cite{franz2012simplified}. Slicer IGT was one of the first software tools to enable easy implementation of image guided intervention software, by integrating existing navigation tools (e.g. the PLUS toolkit --\url{www.plustoolkit.org}, and OpenIGTLink \cite{OpenIGTLink2009}) into Slicer (\url{www.slicer.org}), a general-purpose medical imaging software written in \CC. Slicer IGT is designed as a layer on top of PLUS (which connects and manages data from sensors and image sources) providing a wide, extensible collection of algorithms, and on top of which an application specific GUI and logic can be built. Conveniently, Slicer's functionality can be extended by custom Python-scripted modules. MITK IGT was published later, and followed a similar paradigm: incorporate image guided tools into MITK, a general purpose image processing software. MITK does have a Python module that allows to query data using Python commands. 

As opposed to Slicer IGT and MITK IGT, PRETUS is a minimal software that has no functionality on its own, other than connecting plug-ins and ensuring that they can run concurrently and communicate between them. All the functionality is brought in by plug-ins, that are built as dynamic libraries loaded at run-time. This facilitates a crucial paradigm shift with respect to MITK IGT or Slicer IGT:  instead of aiming at being compatible with the greatest number of devices, PRETUS is conceived to be as independent as possible from specific devices, by delegating most functionality to plug-ins, so that if required a self-contained device specific plug-in can be implemented. 
This design paradigm also promotes that functionality is modular, and that each plug-in does a simple task on a specific input and produces a specific output. Additionally, this allows a very flexible interconnection of plug-ins, for example enabling multiple inputs, outputs, and plug-ins interconnected in arbitrary ways as defined by the user that can be changed during the imaging session.


PRETUS was developed within the iFIND project (\url{www.ifindproject.com}) to collect data and test methods in over 500 pregnant patients. The software has been used for 2D and 3D ultrasound applications. Methods that have used PRETUS in 3D imaging applications include 3D whole-fetus imaging by fast registration of a sequence of 3D ultrasound volumes in real time \cite{Gomez2017FastReg,Gomez2019Manifold},  full placenta imaging by fusion and segmentation of the placenta from multiple 3D ultrasound views \cite{Zimmer2019Placenta,Zimmer2020Placenta}, and whole fetal head imaging using atlas-registration and fusion \cite{Wright2018LSTM,Wright2019Head}. PRETUS has also been used in real time 2D applications, such as standard fetal plane detection\cite{baumgartner2017sononet}, automatic biometric measurements in standard fetal planes \cite{Sinclair2018,Budd2019Head}, and automatic detection and localization of fetal organs from ultrasound images \cite{Toussaint2018Weakly}. PRETUS is also being used in research towards implementing AI-enabled ultrasound methods in low and middle income countries in the context of the VITAL project (\url{http://vital.oucru.org/}), specifically for lung ultrasound in dengue patients \cite{kerdegari2021automatic}.

The software is used via a command-line executable where the user defines a real-time pipeline at run time. The specific experimental setting will depend on the desired pipeline, however a typical setting would be to define an input imaging source (e.g. a file from disk or a framegrabber), a processing task (for example, detecting standard planes) and an output task (e.g. display the results on a screen). More examples and use cases are described in more detail in Sec. \ref{sec:examples}.


In summary, PRETUS is a lightweight, extensible software that addresses the three challenges outlined above as follows: firstly, by enabling the collection of real-time US data from virtually any machine using the video output. Secondly, by enabling real-time pipelines of multiple image processing and visualization steps concurrently. And thirdly, by showing both the live imaging stream, information and outputs from the different processing tasks in a live, compact and unified way. Moreover, PRETUS can take pre-recorded videos or images and play them back at acquisition frame-rate to simulate live sessions in the lab.

\section{Software description}
\label{sec:software_description}

PRETUS is a command-line software, built using open-source software and tested in Linux (all dependencies are cross-platform, but limited testing has been carried out in Windows and Mac) to facilitate research on real-time ultrasound imaging. The software works by establishing a real-time processing pipeline with an arbitrary number of elements (plug-ins). The plug-ins in the pipeline carry out specific functions such as to generate a real-time stream of data (for example from an ultrasound video source), to apply real-time algorithms on the data stream (e.g. implemented via deep neural networks), and to output the result (for example by visualising the processed images, or the metadata, or saving both to a file).

\subsection{Software Architecture}
\label{sec:software_architecture}

\begin{figure}[htb!]
    \centering
    \includegraphics[width=\linewidth]{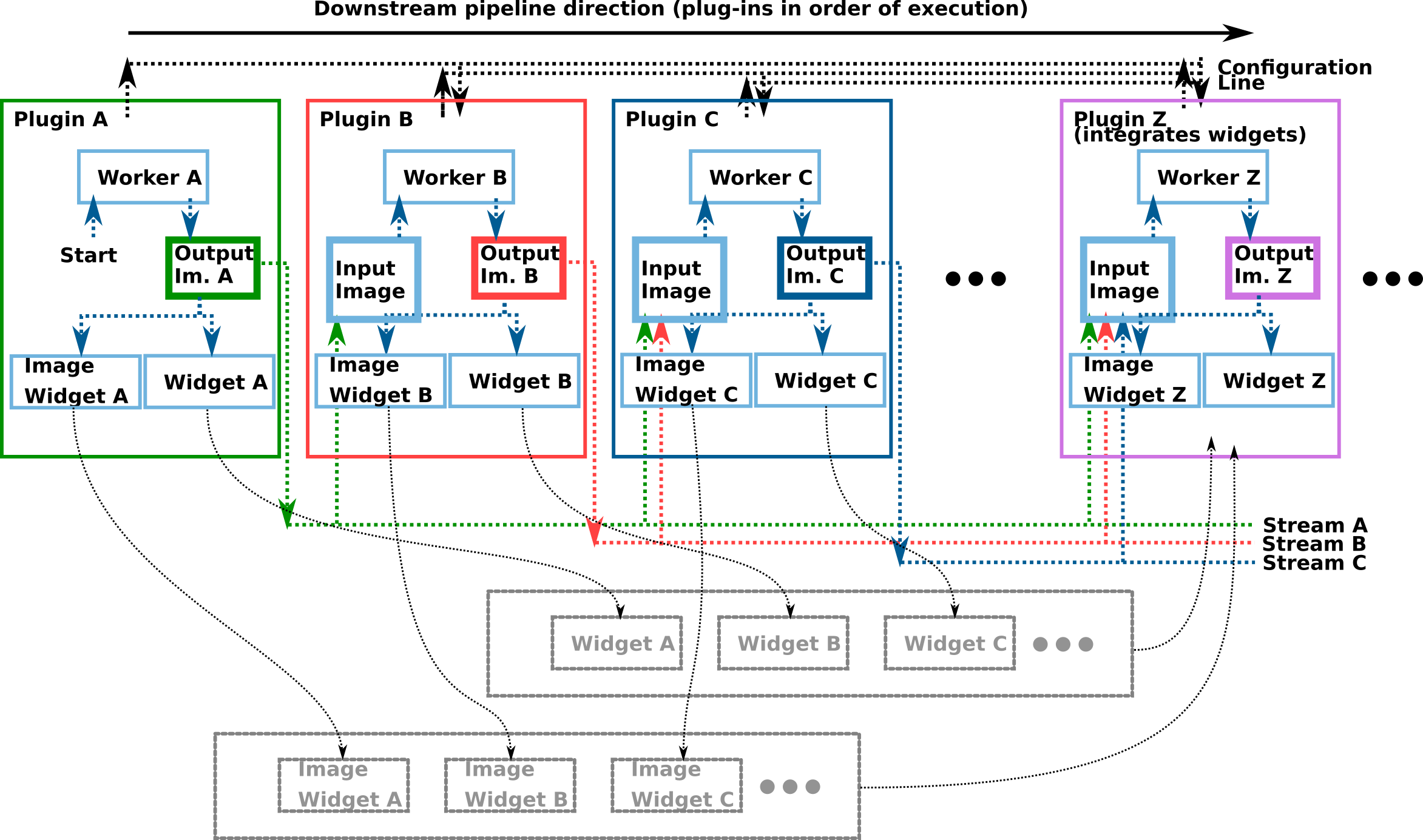}
    \caption{Overview of the plug-in pipeline concept. Each box represents a plug-in, inserted into the execution pipeline in order from left to right. The first plug-in (A) will normally be the imaging source (e.g. frame grabber or file reader) and the last plug-in will normally be for visualization (as illustrated here, integrating widgets from all plug-ins). Configuration is transmitted downstream (illustrated by the configuration line on top) and data is transmitted from each plug-in downstream using the Streams concept (Sec. \ref{sec:streams}). }
    \label{fig:pipeline_overview}
\end{figure}

The software is implemented as a lightweight \texttt{QApplication} (\url{https://www.qt.io/}), which interconnects and starts a number of plug-ins in a pipeline. Figure \ref{fig:pipeline_overview} shows an illustration of the pipeline with the plug-ins and the data transmission lines, or data \emph{Streams} described later in Sec. \ref{sec:streams}.

The software is organised into four modules:
\begin{itemize}
    \item The PRETUS app, in the \texttt{App} folder. The application first loads and instantiates the plug-ins (implemented as dynamic libraries) that are found in the plug-ins folder (or folders) at run-time, also passing any command-line arguments to every plug-in. All plug-ins inherit from the \texttt{Plugin} class, which implements asynchronous callbacks (using QT signals and slots) to transmit configuration information and data between plug-ins. These transmission lines (here referred to as \emph{Streams} and described in more detail in Sec. \ref{sec:streams}) are established, and finally the plug-ins are \emph{activated}, starting the execution loop for the entire plug-in pipeline.
    \item The \texttt{Common} module, which includes common classes, inherited from the iFIND project, to manage data. The main class in this module is the \texttt{ifind::Image} class which is used in PRETUS to encapsulate both images and metadata, and is transmitted through the signal/slots.
    \item The \texttt{PluginLib} library, which implements all classes that plug-ins need to inherit (mainly \texttt{Plugin}, \texttt{Worker} and \texttt{QtPluginWidgetBase}). This library is described in more detail in Sec.~\ref{sec:plugins}.
    \item The \texttt{Plugins} folder, which includes some basic plug-ins readily released with the software. The plug-ins included with this release are further described through examples in Sec. \ref{sec:examples} and in \ref{sec:app:plugins}.
\end{itemize}

Since the main functionality is brought in via the plug-ins, in the following we describe the architecture of the plug-ins in more detail. 

\subsection{Software Functionalities}
\label{sec:software_functionality}

PRETUS defines, at run time, a pipeline of imaging plug-ins that in sequence process a stream of images. The sources of the imaging data, the processes themselves and whether the outcomes are displayed and/ or stored depends on the plug-ins used. In terms of performance, PRETUS is designed to satisfy two main requirements. First, plug-ins can be executed concurrently, i.e., the work of each plug-in runs in a separate non-blocking thread. Second, plug-ins must run as close as possible to real-time. The work of each plug-in runs at a user-defined frame rate, and the latest available frame is processed when a previous computations has been completed. To this end, PRETUS will drop frames at the input of a plug-in until the plug-in is ready to accept a new one. This behaviour can be overridden if a specific plug-in does not need real-time performance and processing of all frames is sought.

The two above functionalities are implemented through two mechanisms: the plug-in system, and the \emph{Streams}.

\subsubsection{Plug-in System}
\label{sec:plugins}

Plug-ins are independent programs, built as dynamic libraries, and are loaded into PRETUS at run time. All plug-ins take images as input, yield images at the output, and normally delegate the processing task to a \texttt{Worker} class. Plug-ins can also have two complementary means of displaying information and outputs: (i) by implementing a widget that will typically show graphs, numbers and text, and allow for input through sliders and other widgets; and (ii) by implementing an image widget that will display images, overlays, masks, etc. Both the widget and the image widget must inherit from the \texttt{QtPluginWidgetBase} class in \texttt{PluginLib}. Two examples of how to build plug-ins for PRETUS are outlined in Sec. \ref{sec:examples} and detailed in the repository (\url{https://github.com/gomezalberto/pretus}).

The basic operation of a plug-in in the pipeline is as follows:
\begin{enumerate}
    \item The plug-in receives an input image from previous plug-ins in the pipeline. The image is also passed on to the next plug-in.
    \item If the image belongs to the Stream(s) that this plug-in accepts, the image is sent to the plug-in's timer.
    \item If the Worker is not processing the previous image, the timer sends the latest image to the worker in a separate thread. \item The main processing task of the plug-in is carried out in the Worker. When finished, the output image is sent to the plug-in and the timer is notified that the worker is ready to take a new image.
    \item The plug-in sends the output image through the plug-in's output Stream. Downstream plug-ins are now able to use it.
    \item If the plug-in has a \texttt{Widget} and/or an \texttt{ImageWidget}, the output image is sent to them for display. The user can act on any inputs available in the widget (e.g. sliders, checkboxes, etc) to make changes in the plug-in behaviour during the imaging session.
\end{enumerate}

We recommend (as we do in all plug-ins included in this repository) that any output image resulting from a plug-in's task is added as a layer to the input image. Because images are transmitted as pointers, no data will be duplicated in memory, so this mechanism is efficient. Additionally, this allows to always track what image was used to produce what result, even if different Streams operate at different rates and while maintaining real-time performance.

\subsubsection{Streams}
\label{sec:streams}
In this context, a \emph{Stream} refers to every image sequence produced by a plug-in and accessible to all other downstream plug-ins in the pipeline. Every stream is named after the plug-in that generates it, with the exception of the plug-ins that generate data at the source, also called input plug-ins.

Input plug-ins capture imaging data and transmit it downstream the pipeline. In the current release of PRETUS, three input plug-ins are provided: the Video Manager plug-in, that can read and transmit frames from a video file; the Frame Grabber plug-in, that reads video output from an ultrasound system and transmits it frame by frame; and the File Manager, that reads images from a folder system and transmits them at a given framerate. These type of plug-ins must return \texttt{true} via the \texttt{IsInput()} plug-in method. The Stream transmitted by an input plug-in is called `Input' regardless of the plug-in name. Multiple input plug-ins can be used simultaneously, in which case only the first will have an Stream called `Input', and the rest (in order of appearance in the pipeline) will be called `Input1', `Input2', etc. The rest of the plug-ins in the pipeline will by default accept images from `Input' but can be set to use other inputs with the command-line option \texttt{--<pluginname>\_stream Input1} or using the menu in the widget during the imaging session.


\subsubsection{Building a plug-in}

Plug-ins are dynamic libraries written in \CC, that link against the \texttt{Plugin} library provided with PRETUS. The processing task carried out by the plug-in can be implemented in \CC, or in Python. To illustrate the two types of plug-ins, the repository includes two sample plug-ins in the \texttt{Plugins} folder designed and documented to serve as templates and tutorials for developers to implement their \CC plug-ins (\texttt{Plugin\_CppAlgorithm}) and their Python plug-ins (\texttt{Plugin\_PythonAlgorithm}).


\section{Illustrative Examples}
\label{sec:examples}



\begin{figure}[htb!]
    \centering
    \begin{subfigure}[t]{0.32\linewidth}
         \centering
         \includegraphics[width=\linewidth]{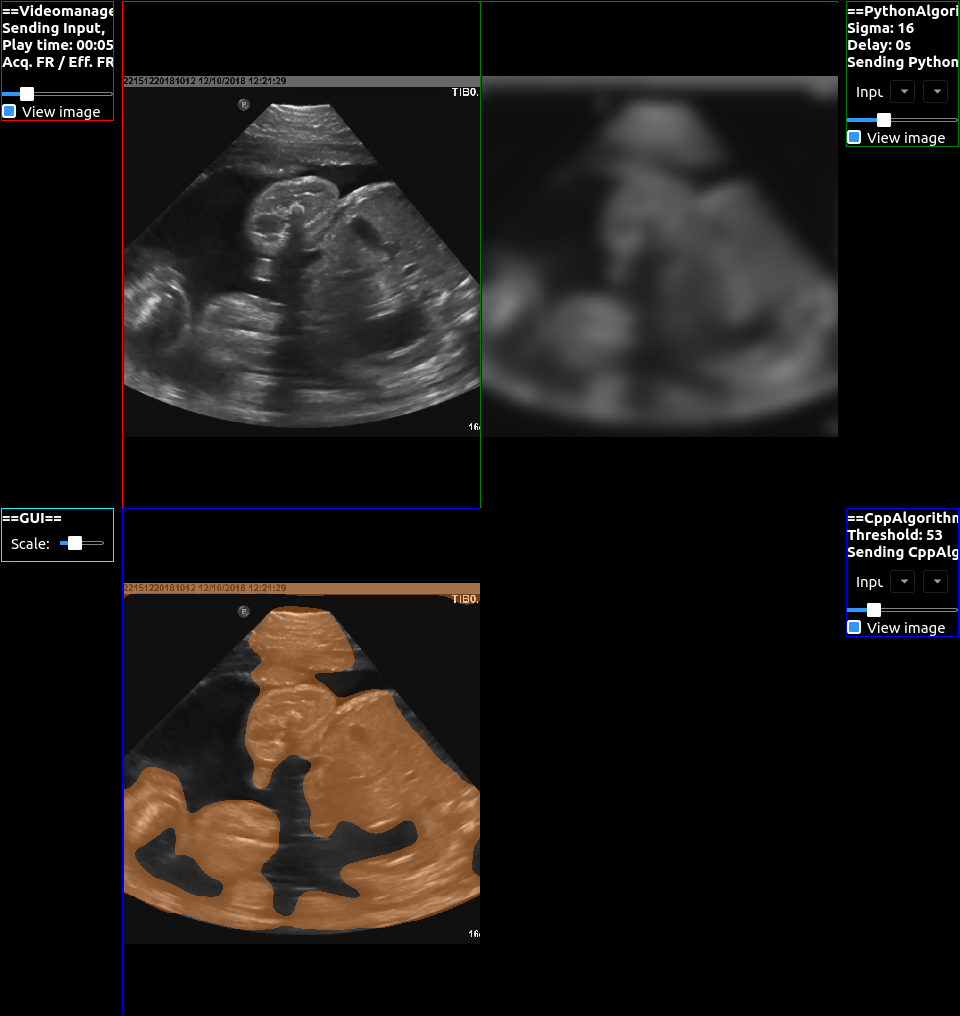}
         \caption{Thresholding and blurring. }
         \label{fig:plugins:blur-threshold}
     \end{subfigure}
     \hfill
    \begin{subfigure}[t]{0.32\linewidth}
         \centering
         \includegraphics[width=\linewidth]{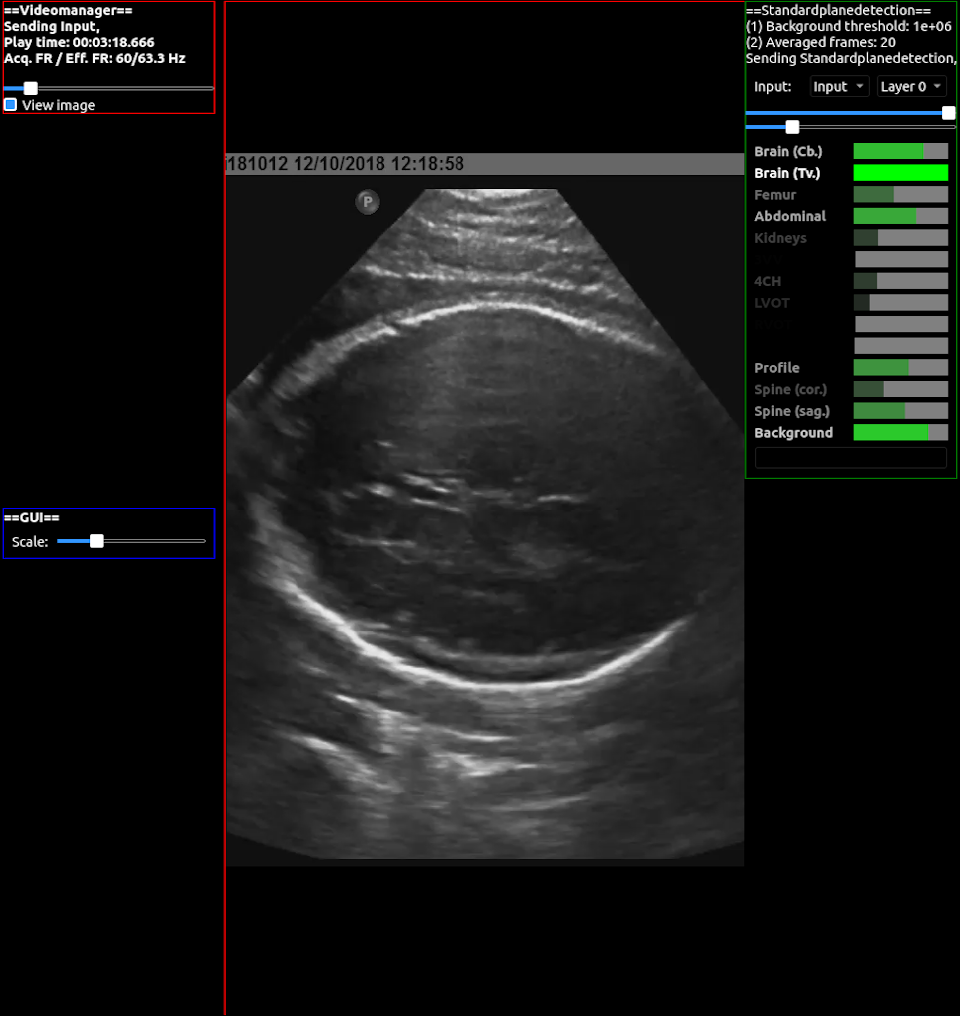}
         \caption{Standard plane detection. }
         \label{fig:plugins:sononet}
     \end{subfigure}
     \hfill
    \begin{subfigure}[t]{0.32\linewidth}
         \centering
         \includegraphics[width=\linewidth]{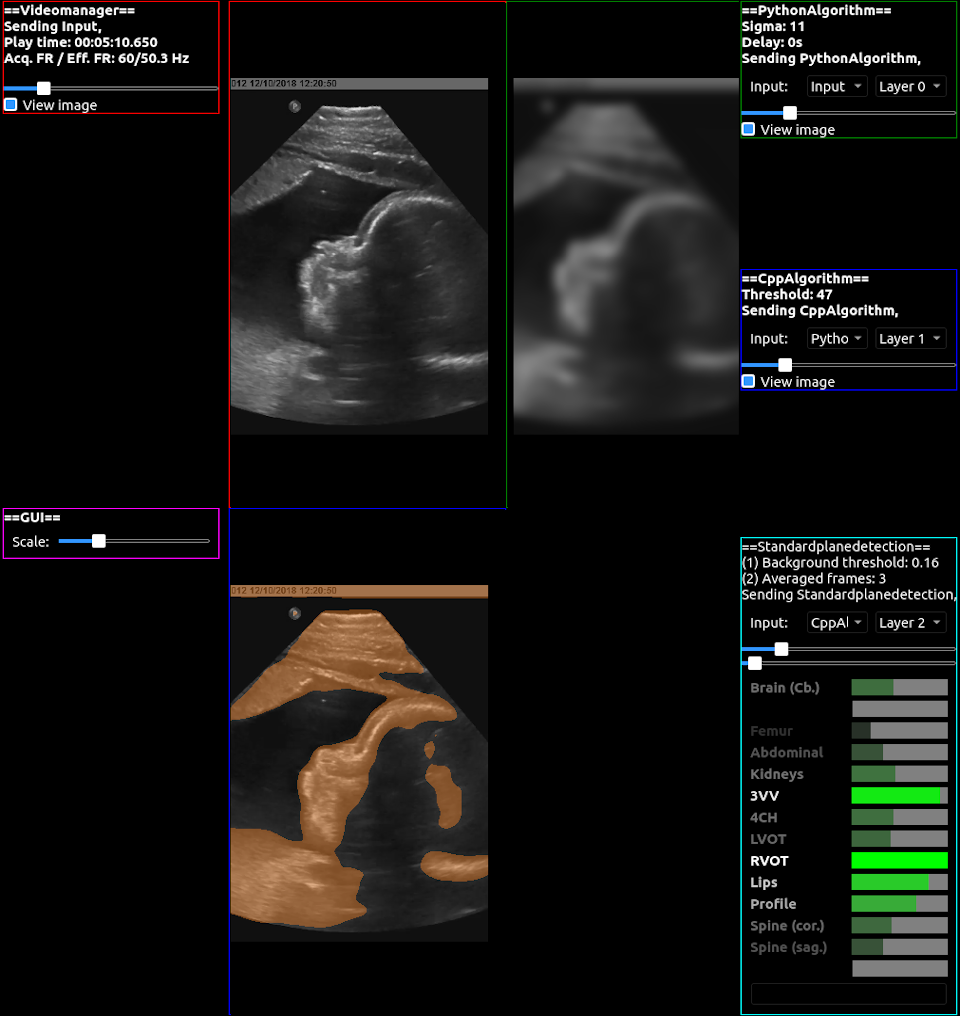}
         \caption{Three concurrent plug-ins.}
         \label{fig:plugins:sononet}
     \end{subfigure}
     \hfill
    \caption{Screenshot of three PRETUS pipelines. (a) pipeline where the input image is shown in the top left, the blurred image on the top right, and the thresholded version of the blurred image, overlaid onto the input, on the bottom left. (b) capture of the Standard Plane Detection plug-in integrating SonoNet in PRETUS where a `Head' standard view has been detected. (c) three plug-ins (blurring, thresholding and Sononet) working concurrently. All examples are described in Sec. \ref{sec:examples}, and illustrated in the videos in the supplementary material. }
    \label{fig:plugins}
\end{figure}

In this section we describe three usage examples: first, an example of real-time blurring and thresholding of an ultrasound video. These plug-ins are not designed with an intended application in mind other than exemplifying the design and performance of PRETUS. Second, an example showing the integration of SonoNet~\cite{baumgartner2017sononet}, a deep neural network for the automatic identification of anatomical fetal standard view planes, into PRETUS. And third, blurring, thresholding and SonoNet working in the same pipeline, where we evaluate the real-time performance with concurrent \CC and Python plugins. Videos showing the three examples in action is provided in the supplementary material, and a screen shot of these videos is shown in Fig. \ref{fig:plugins}.

\subsection{Blurring and Thresholding}

The purpose of this example is to illustrate the connection of the output of a plug-in to the input of another plug-in, and the visualization of the results. To this end, we build a pipeline with four plugins: the video manager plug-in (Sec. \ref{sec:plugin:videomanager}) as the input plug-in, which reads a video from file and transmits each frame through the pipeline. Then, the Python Algorithm plug-in (Sec. \ref{sec:plugin:python}) takes the video frames and applies a blurring operation. The blurred frame is input to the Cpp Algorithm plug-in (Sec. \ref{sec:plugin:cpp}), which performs a binary thresholding operation on the blurred image. Finally, the GUI plug-in (Sec. \ref{sec:plugin:gui}) takes all widgets from the three previous plug-ins and displays them on screen.

All plug-ins accept, by default, images from the \texttt{Input} stream. This stream is generated by any of the input plug-ins (video manager, frame grabber and file manager) which are first in the pipeline. For the Cpp Algorithm plug-in to receive the output of the Python Algorithm plug-in as input, we use the optional argument \texttt{--cppalgorithm\_stream pythonalgorithm} (or select the input from the Python Algorithm, and the last layer, in the widget). In addition, the Python Algorithm adds the blurred image as an additional layer to its input image, and the Cpp Algorithm plug-in needs to be informed of what layer from the \texttt{pythonalgorithm} stream to use, in this case the last one, with the command line argument \texttt{--cppalgorithm\_layer -1}. The complete command line call for this example is:
\begin{lstlisting}[language=bash,
basicstyle=\fontfamily{pcr}\selectfont\scriptsize\color{white},
                    backgroundcolor=\color{darkgray},
                    frame=single,]
$ ./bin/pretus -pipeline "videomanager>pythonalgorithm>cppalgorithm>gui" \ 
  --videomanager_input ~/data/video.MP4 \ 
  --cppalgorithm_stream pythonalgorithm \ 
  --cppalgorithm_layer -1
\end{lstlisting}
The program will launch, display information about the plug-ins used as below, and open a window with the GUI (Fig. \ref{fig:plugins}, left, and Video SV1).

\begin{lstlisting}[language=bash,
                    basicstyle=\fontfamily{pcr}\selectfont\tiny\color{white},
                    backgroundcolor=\color{darkgray},
                    frame=single,]
Loading plug-ins from <folder>
  0 [Plugin] loading <folder>/libPlugin_filemanager.so...	File manager(0) loaded
  1 [Plugin] loading <folder>/libPlugin_imageFileWriter.so...	Image file writer(1) loaded
  2 [Plugin] loading <folder>/libPlugin_videomanager.so...	Video manager(2) loaded
  3 [Plugin] loading <folder>/libPlugin_CppAlgorithm.so...	Cpp Algorithm(3) loaded
  4 [Plugin] loading <folder>/libPlugin_framegrabber.so...	Frame grabber(4) loaded
  5 [Plugin] loading <folder>/libPlugin_PythonAlgorithm.so...	Python Algorithm(5) loaded
  6 [Plugin] loading <folder>/libPlugin_planeDetection.so...	Standard plane detection(6) loaded
  7 [Plugin] loading <folder>/libPlugin_GUI.so...	GUI(7) loaded
Video manager -> Python Algorithm
Python Algorithm -> Cpp Algorithm
Cpp Algorithm -> GUI
VideoManager::Initialize() - loading video ~/data/video.MP4... loaded, FPS = 60, frames = 110842
Start acquisition
Manager::exitLoop() - Enter 'quit' to exit:
>>
\end{lstlisting}
The program will exit by entering `quit' in the command line.

\subsection{SonoNet integration}
In this example we illustrate the integration of SonoNet \cite{baumgartner2017sononet}, a model to detect standard fetal planes for the 20 week fetal screening ultrasound examination. The SonoNet model is incorporated into pretus via the Standard Plane Detection plug-in.

Since SonoNet is implemented in a frame by frame basis, in our implementation we allow the user to use a temporal average to leverage high acquisition frame rates to stabilise the plane prediction. The number of frames to be averaged can be set by command line argument and modified in real time via a slider in the widget. With this, the resulting call is:
\begin{lstlisting}[language=bash,
basicstyle=\fontfamily{pcr}\selectfont\scriptsize\color{white},
                    backgroundcolor=\color{darkgray},
                    frame=single,]
$ ./bin/pretus -pipeline "videomanager>standardplanedetection>gui" \ 
  --videomanager_input ~/data/video.MP4 --standardplanedetection_taverage 20
\end{lstlisting}
The resulting display and interactions can be seen in Fig. \ref{fig:plugins}, middle, and Video SV2.

\subsection{All plug-ins in the same pipeline}

The purpose of this example is to demonstrate that multiple \CC and Python plug-ins can work concurrently, and to evaluate to the effect of delay and execution time of each plug-in in the performance of the entire pipeline and the overall delay with respect to the input stream. We use the Video Manager plug-in as input, and use the Python Algorithm, the Cpp Algorithm and SonoNet in the pipeline, followed by the GUI plug-in. To illustrate the behaviour of frame-dropping at high delays, we introduce an artificial variable wait time in the Python Algorithm plug-in. All plug-ins are provided with the command-line option \texttt{<pluginname>\_time 1}, which measures the execution time of the worker. We run the pipeline in five configurations: 1) with all plug-ins using the \texttt{Input} stream; and 2) to 5), with plug-ins connected in sequence, using as input the output stream of the previous plug-in, and a plug-in frame rate (identical for all three plug-ins) of 10Hz (configuration 2), 20Hz (configuration 3), 30Hz (configuration 4), and 40Hz (configuration 5). We also measure execution times for all plug-in. The execution call for the first configuration is:
\begin{lstlisting}[language=bash,
basicstyle=\fontfamily{pcr}\selectfont\tiny\color{white},
                    backgroundcolor=\color{darkgray},
                    frame=single,]
$ ./bin/pretus -pipeline "ideomanager>pythonalgorithm>cppalgorithm>standardpanedetection>gui" \ 
--videomanager_input ~/data/video.MP4 \ 
--standardplanedetection_time 1 --pythonalgorithm_time 1 --cppalgorithm_time 1 \ 
 --pythonalgorithm_delay 0.1
\end{lstlisting}
For configurations 2) to 5) (replacing the frame rate value):
\begin{lstlisting}[language=bash,
basicstyle=\fontfamily{pcr}\selectfont\tiny\color{white},
                    backgroundcolor=\color{darkgray},
                    frame=single,]
$ ./bin/pretus -pipeline "videomanager>pythonalgorithm>cppalgorithm>standardpanedetection>gui" \ 
--videomanager_input ~/data/video.MP4 --videomanager_verbose 1 \ 
--standardplanedetection_time 1 --pythonalgorithm_time 1 --cppalgorithm_time 1 \ 
--pythonalgorithm_framerate 10 --pythonalgorithm_delay 0.1 \ 
--cppalgorithm_framerate 10 --standardplanedetection_framerate 10 \
--cppalgorithm_stream pythonalgorithm --cppalgorithm_layer -1 \ 
--standardplanedetection_stream cppalgorithm  --standardplanedetection_layer -1
\end{lstlisting}
An example showing the visualization for configuration 2 can be seen in Fig. \ref{fig:plugins}, right, and in Video SV3. The table \ref{tab:time-results} shows the average $\pm$ standard deviation execution time per plug-in, for different wait times introduced in the first plug-in in the sequence (the Python Algorithm plug-in -PA).  These wait times (from 0 to 200 ms, as indicated in the table) have no effect in the execution time of other plug-ins downstream because each is executed on a separate thread.

\begin{table}[!htb]
\begin{tabular}{|c|c|c|c|c|c|c|}
\hline
& Wait & 0ms & 50ms & 100ms & 150ms & 200ms \\ \hline 
\multirow{3}{*}{Time (ms)} & PA & 26.2$\pm$5 & 76.7$\pm$7 & 124.4$\pm$3 & 173.6$\pm$1 & 223.9$\pm$1 \\ 
& CA & 0.9$\pm$1 & 0.9$\pm$1 & 0.8$\pm$1 & 0.8$\pm$0 & 0.7$\pm$1 \\ 
& SPD & 17.6$\pm$6 & 17.7$\pm$9 & 15.0$\pm$3 & 14.6$\pm$1 & 14.8$\pm$2 \\  
\hline
\end{tabular}
\caption{Average $\pm$ standard deviation of the execution time, in ms
, of each plug-in: Python Algorithm (PA), Cpp Algorithm (CA) and Standard plane detection (SPD), in configuration 1).}
\label{tab:time-results} 
\end{table}

\begin{figure}[htb!]
    \centering
    \includegraphics[width=\linewidth]{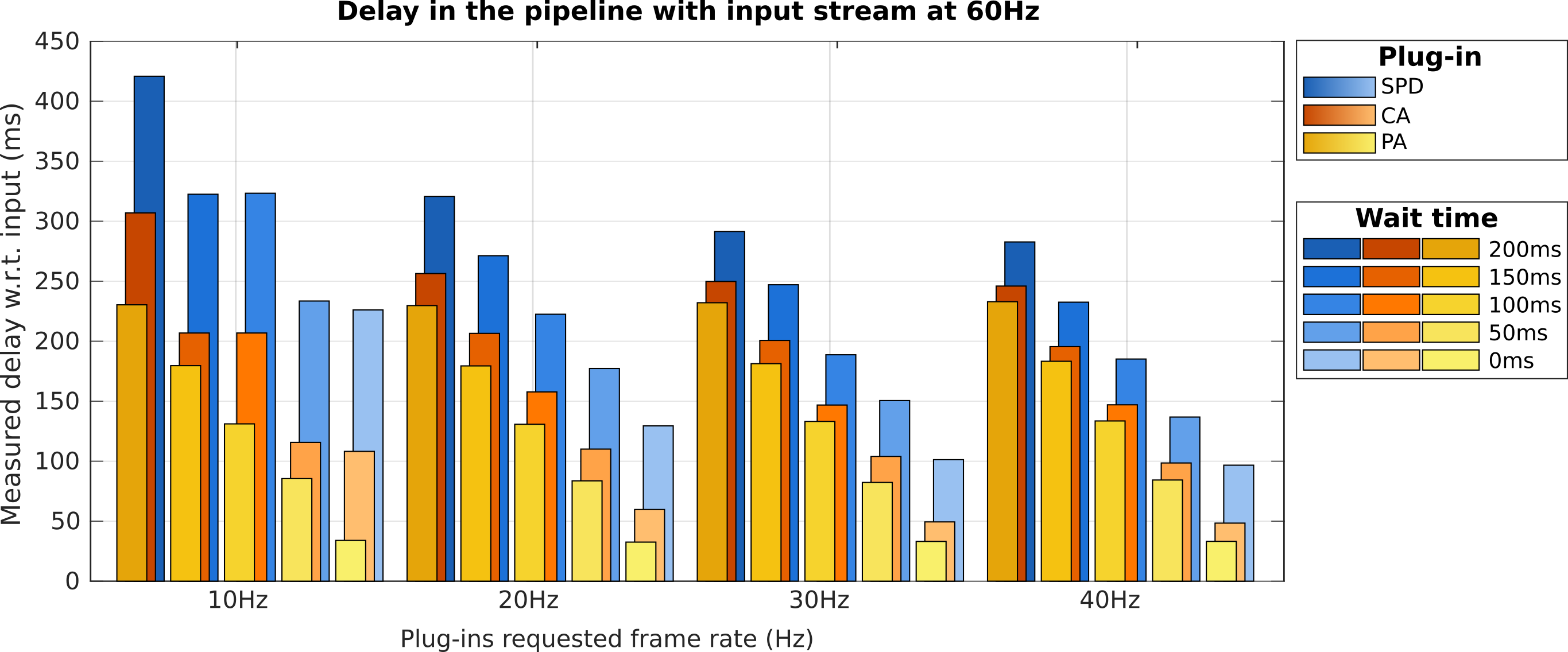}
    \caption{Measured delay between the output of each plug-in and real-time input stream when plug-ins are connected in sequence: Input $\rightarrow$ Python Algorithm (PA, in yellow) $\rightarrow$ Cpp Algorithm (CA, in orange) $\rightarrow$ SonoNet (SPD, in blue). The increasing wait time introduced in the PA plug-in is coded in the lighter to darker shade for each plug-in. When a higher frame-rate is requested, the delay decreases until around 30Hz where other system parameters (e.g. memory, CPU, etc.) become the limiting factor. }
    \label{fig:delay_vs_fr}
\end{figure}

When the plug-ins are set up in sequence (accepting the input from the previous plug-in, configurations 2) to 5)), their individual execution time are not affected. As shown in Table \ref{tab:time-results} the decrease in execution time observed in the Standard Plane Detection plug-in as the wait time increases is due to more CPU resource available which is used in the pre-processing steps and other CPU based tasks. However each plug-in still needs to wait to receive the image from the previous plug-in, which introduces an added delay compared to processing images from the `Input' stream. We measured the total delay between the output of a plug-in and the `Input' stream by tracking the frame number from the input, and comparing the timestamp of that frame after processing. This delay will vary depending on the requested plug-in frame rate. Plug-ins check for the latest input frame at this requested frame rate (20Hz by default), and new frames are dropped until the plug-in has finished the current processing task to avoid temporal drift. When idle, the next image is processed when the next periodic input check arrives. These two effects are illustrated in Fig. \ref{fig:delay_vs_fr}.

Because of the frame-dropping, the effective frame rate, i.e. the number of frames per second that each plug-in actually processes depends on three factors: 1) the execution time of the plug-in (which limits the maximum effective frame rate), 2) the execution time of the plug-ins that precede the current plug-in, if connected in sequence (the current plug-in will have to wait), and 4) other system parameters (e.g. memory, CPU, etc). To illustrate the effect of these factors, we report the measured frame rates for each plug-in in Table~\ref{tab:fr-results}. 

\begin{table}[!htb]
\begin{tabular}{|c|c|c|c|c|c|c|}
\hline
Conf. & Wait & 0ms & 50ms & 100ms & 150ms & 200ms \\ \hline  \hline
\multirow{3}{*}{\makecell{1)\\ Par.\\ 20Hz}} & PA &\cellcolor{gray!20}  20.0$\pm$2.5 & 9.8$\pm$1.0 & 6.6$\pm$0.5 & 4.9$\pm$0.3 & 3.9$\pm$0.2 \\ 
& CA &\cellcolor{gray!20}  20.0$\pm$2.1  &\cellcolor{gray!20}  20.3$\pm$3.4  &\cellcolor{gray!20}  20.4$\pm$3.5  &\cellcolor{gray!20}  20.3$\pm$3.7  &\cellcolor{gray!20}  20.3$\pm$2.9  \\ 
& SPD &\cellcolor{gray!20}  20.2$\pm$3.1  &\cellcolor{gray!20}  19.9$\pm$4.0  &\cellcolor{gray!20}  20.0$\pm$4.4  &\cellcolor{gray!20}  20.0$\pm$4.2  &\cellcolor{gray!20}  20.0$\pm$3.3  \\ 
\hline \hline
\multirow{3}{*}{\makecell{2)\\ Seq.\\ 10Hz}} &  PA & \cellcolor{gray!20} 10.2$\pm$1.5 & \cellcolor{gray!20} 9.6$\pm$1.4 & 5.0$\pm$0.1 & 5.0$\pm$0.1 & 3.3$\pm$0.6 \\ 
& CA & \cellcolor{gray!20} 10.0$\pm$0.7  & \cellcolor{gray!20} 9.7$\pm$1.4  & 5.0$\pm$0.1  & 5.0$\pm$0.1  & 3.3$\pm$0.5  \\ 
& SPD & \cellcolor{gray!20} 10.3$\pm$1.9  & \cellcolor{gray!20} 9.8$\pm$1.8  & 5.1$\pm$0.6  & 5.0$\pm$0.4  & 3.3$\pm$0.6  \\ 
\hline \hline
\multirow{3}{*}{\makecell{3)\\ Seq.\\ 20Hz}} & PA &  \cellcolor{gray!20} 19.6$\pm$2.6 & 9.9$\pm$0.8 & 6.7$\pm$0.3 & 5.0$\pm$0.1 & 4.0$\pm$0.1 \\ 
& CA &  \cellcolor{gray!20} 19.6$\pm$2.5  & 9.9$\pm$0.8  & 6.7$\pm$0.3  & 5.0$\pm$0.1  & 4.0$\pm$0.0  \\ 
& SPD &  \cellcolor{gray!20} 19.8$\pm$3.2  & 10.0$\pm$1.4  & 6.7$\pm$0.5  & 5.0$\pm$0.2  & 4.0$\pm$0.1  \\ 
\hline \hline
\multirow{3}{*}{\makecell{4)\\ Seq.\\ 30Hz}} & PA & \cellcolor{gray} \textcolor{white}{25.9$\pm$5.0} & 10.6$\pm$0.9 & 7.4$\pm$0.5 & 5.2$\pm$0.1 & 4.1$\pm$0.6 \\ 
& CA & \cellcolor{gray} \textcolor{white}{ 26.0$\pm$5.1}  & 10.5$\pm$0.9  & 7.4$\pm$0.5  & 5.2$\pm$0.1  & 4.1$\pm$0.6  \\ 
& SPD & \cellcolor{gray} \textcolor{white}{ 27.1$\pm$7.9 } & 10.8$\pm$2.2  & 7.5$\pm$1.1  & 5.2$\pm$0.3  & 4.1$\pm$0.6  \\ 
\hline \hline
\multirow{3}{*}{\makecell{5)\\ Seq.\\ 40Hz}} & PA & \cellcolor{gray} \textcolor{white}{26.3$\pm$4.0} & 11.4$\pm$1.1 & 7.3$\pm$0.4 & 5.4$\pm$0.2 & 4.2$\pm$0.1 \\ 
& CA & \cellcolor{gray} \textcolor{white}{26.4$\pm$4.3}  & 11.4$\pm$1.2  & 7.3$\pm$0.4  & 5.4$\pm$0.2  & 4.2$\pm$0.1  \\ 
& SPD & \cellcolor{gray} \textcolor{white}{27.6$\pm$7.9 } & 11.7$\pm$2.1  & 7.5$\pm$1.7  & 5.4$\pm$0.4  & 4.2$\pm$0.2  \\ 
\hline
\end{tabular}
\caption{Average $\pm$ standard deviation of the measured effective frame rate, in Hz, of each plug-in: Python Algorithm (PA), Cpp Algorithm (CA) and Standard plane detection (SPD), when the three plug-ins are executed in parallel (Par. row) or connected in sequence (Seq. rows) at the indicated user-requested frame rates (the same for all three plug-ins). Cells highlighted in light gray indicate that the measured and the requested frame rate match. }
\label{tab:fr-results} 
\end{table}

As expected, in parallel (Par.), where all plug-ins use data from the \texttt{Input} stream, the CA and SPD plug-ins maintain the requested frame rate (20Hz) independently of the other plug-ins. Obviously, the PA plug-in can only maintain the frame rate when no wait time is introduced, and then the frame rate decreases inversely to the wait time. This further demonstrates that two plug-ins, where the task is implemented in Python, can run independently in parallel. This is achieved by sharing the Python interpreter across plug-ins. In sequential execution (Seq.), the frame rate is limited by the wait time; as a result, the requested frame rate can only be achieved in certain cases (highlighted in light grey in the table). When requesting very high framerates, the pipeline might not be able to deliver at that framerate and the framerate will be capped to the maximum system framerate for that specific pipeline, which is around 26Hz in this case (highlighted in dark gray in Table \ref{tab:fr-results}). 

\section{Impact}
\label{sec:impact}


PRETUS will promote and facilitate real-time ultrasound imaging research for two main reasons: first, PRETUS is plug-in based, and plug-ins are self-contained in the sense that they implement the data processing, argument handling, user interface, image visualization, and any other display or input widgets; however, default modules and basic building blocks to develop plug-ins are provided in the \texttt{Plugin} library, and examples of plug-ins using both \CC and Python are provided, simplifying the implementation of new plug-ins. Second, PRETUS implements user-transparent, multi-threaded plug-in execution with periodic calls to the tasks implemented by the plug-ins, to ensure that plug-ins always use the latest generated input image and that they run as close to real-time as possible. Crucially, a Python interpreter is shared across Python plug-ins enabling concurrent execution of independent, dynamically loaded Python plug-ins too.


Indeed, hundreds of research papers on ultrasound image analysis are published every year, most of which are trained and tested offline using an image database. As a result a platform to facilitate real-time data collection and implementing and evaluating computational methods in a realistic, real-time clinical scenario connected to an ultrasound imaging system is highly sought in US research. PRETUS allows the integration of \CC and Python methods in a simple and flexible way, with minimal changes to an offline version that users and developers may already have. To this end, we have included tutorials on how to build plug-ins both in \CC and Python. Conveniently, PRETUS also allows playing-back captured videos or sequences of images retrospectively and at acquisition frame rates, to replicate live sessions offline, in the lab. Being a lightweight and plug-in based software, PRETUS can accelerate translational research in ultrasound imaging for diagnostic and interventions.


Unlike other software, PRETUS implements real-time execution transparently to users and plug-in developers, by ensuring that, regardless of the performance and speed of the computational method, the latest input image will always be fed to the algorithm to avoid execution drift. PRETUS also ensures that plug-ins run in parallel and that their outputs and inputs can be interconnected at run time. A unique feature of PRETUS is that plug-ins can be implemented in Python and in \CC and multiple Python plug-ins can run concurrently by sharing the Python interpreter. Moreover, PRETUS is designed to connect to the video output of virtually any ultrasound system, so as to remove the impediment of manufacturer-specific formats and transmission protocols. However, a developer can easily implement a machine specific acquisition plug-in, using the provided input plug-ins as examples, if machine specific protocols are made available.


PRETUS also includes a file saving plug-in which turns the system into a powerful data collection software that can facilitate the acquisition of large amounts research data. Because multiple Input Streams can be captured simultaneously, synchronised multi-source data can also be captured and stored. Additionally, plug-in developers may implement annotation plug-ins for live image annotation, allowing not only to collect data but to annotate it rapidly at collection time.




\section{Conclusions}
\label{sec:conclusions}

We have discussed PRETUS, a plug-in based, real-time software for US imaging research. The software allows the collection of live ultrasound data and the use of algorithms (implemented in \CC or Python). PRETUS loads plug-ins dynamically and a plug-in pipeline can be defined by the user at run time.

We have evaluated PRETUS with three examples, demonstrating real-time, concurrent execution of multiple plug-ins. PRETUS can be extended easily through more plug-ins and has the potential to enable researchers to evaluate their methods in real-time with minimal implementation efforts.

\section{Conflict of Interest}



We wish to confirm that there are no known conflicts of interest associated with this publication and there has been no significant financial support for this work that could have influenced its outcome.






\section*{Acknowledgements}

This work was supported by the Wellcome Trust IEH Award [102431], by the Wellcome/EPSRC Centre for Medical Engineering [WT203148/Z/16/Z] and by the National Institute for Health Research (NIHR) Biomedical Research Centre at Guy's and St Thomas' NHS Foundation Trust and King's College London. The views expressed are those of the author(s) and not necessarily those of the NHS, the NIHR or the Department of Health.

\appendix

\section{Plug-ins included with this release}
\label{sec:app:plugins}

\subsection{File manager plug-in}
\label{sec:plugin:filemanager}

This plug-in allows images to be read from a sub-directory hierarchy and transmits them through the pipeline at a certain frame-rate. Images can be 2D or 3D, and the mhd/raw format from the ITK library is preferred. Other formats supported by ITK can be also used by changing the expected file extension with the \texttt{-filemanager\_extension} command line argument.

By default, images are transmitted in alphabetical order, therefore the file name will dictate the transmission order. Also, by default, images are transmitted at a constant frame rate of 20 images per second. A custom frame rate can be set by the user, in two ways: first, a constant frame rate between 0 and 200 can be set using the command line argument \texttt{-filemanager\_framerate}. Second, if the mhd headers have the field \texttt{AcquisitionFrameRate}, then this value will be used, and can be different for each image. Additional options allow the last image to loop around when it is read or to ignore the header information.

\subsection{Video manager plug-in}
\label{sec:plugin:videomanager}

This plug-in allows a video file to be read from the file system and transmits it through the pipeline. \texttt{Opencv} is used to read the video files so supported format depends on local configuration of opencv.

The video by default loops around when finished, but this can be disabled by the user using the command line argument \texttt{-videomanager\_loop 0}. The video starts from the beginning by default, but an arbitrary start time can be set with \texttt{-videomanager\_start\_time <mm:ss>}. The video can also be played faster by setting a fast-forward factor with \texttt{-videomanager\_ff <factor>}. This plug-in also enables interactively moving around in the video with a slider in the plug-in's widget.

\subsection{Frame grabber plug-in}
\label{sec:plugin:framegrabber}

This plug-in allows a stream of images to be received in real-time from a video source, such as the video output of an ultrasound system, by using the Epiphone DVI2USB3.0 frame grabber (\url{https://www.epiphan.com/products/dvi2usb-3-0/}). The plug-in is currently implemented to convert the images to grayscale and pass it on to the rest of the pipeline as a single channel, 8 bit images.

\subsection{Cpp Algorithm plug-in}
\label{sec:plugin:cpp}

This plug-in performs a simple binary thresholding on the input image. The plug-in is conceived as a tutorial to illustrate how to develop \CC plug-ins for PRETUS.

The Cpp Algorithm plug-in performs the thresholding operation using the ITK library. The threshold value can be set  via command-line argument (\texttt{cppalgorithm\_th <th>}) and edited in real-time using the slider in the plug-in's widget. An overlay of the input image and the thresholded image are shown on the plug-in's image widget.

\subsection{Python Algorithm plug-in}
\label{sec:plugin:python}

This plug-in performs a Gaussian blur on the input image. The plug-in is conceived as a tutorial to illustrate how to develop Python plug-ins for PRETUS.

The Python Algorithm plug-in performs the Gaussian blur operation using the SimpleITK Python library. The sigma value for the Gaussian kernel can be set  via command-line argument (\texttt{pythonalgorithm\_sigma <sigma>}) and edited in real time using the slider in the plug-in's widget. The blurred version of the input image is shown on the plug-in's image widget. The plug-in's worker waits a user-defined time (within the Python code) to simulate a longer task execution.

\subsection{Standard plane detection (SonoNet)}
\label{sec:plugin:sononet}

This plug-in implements the fetal scan plane detection method described in~\cite{baumgartner2017sononet}. The plug-in runs the method in every frame received from the input stream (which can be selected by the user). 

The model makes a prediction about the scan plane corresponding to the image, and classifies the image into one of 13 standard views: `3VV' (cardiac three vessel view), `4CH' (cardiac four chamber), `RVOT' (cardiac right ventricular outflow tract), `LVOT' (cardiac left ventricular outflow tract), `Abdominal', `Brain (Cb.)' (cerebellum), `Brain (Tv.)' (trans-ventricular), `Femur', `Kidneys', `Lips', `Profile', `Spine (cor.)' (coronal), `Spine (sag.)' (sagittal), or `Background'. Illustrative examples of these views and their significance can be found in \cite{nhsscreening}. 

The algorithm yields a 13-element vector with a score indicating the probability of the image belonging to each class above. The plug-in packs this information into four fields in the output image header: 
\begin{itemize}
    \item ``\texttt{Standardplanedetection\_labels}'', a string array with the original class labels in order.
    \item ``\texttt{Standardplanedetection\_confidences}'', a float array with the probability for each class. 
    \item ``\texttt{Standardplanedetection\_label}'', a string with the label of the highest scoring class
    \item ``\texttt{Standardplanedetection\_confidence}'', a float with the probability of the highest scoring class.
\end{itemize}

This output image is transmitted downstream the pipline in the \texttt{Standardplanedetection} stream. The visualization widget displays these information as a bar plot with the classes and probabilities.

\subsection{Image file writer plug-in}
\label{sec:plugin:filewriter}

This plug-in allows images to be written to file, in real-time. The plug-in can write images from any stream, or multiple Streams, or all. 

This plug-in handles the header field ``DO\_NOT\_WRITE'' by not writing to file any image that has that key in the header, even if the image belongs to a stream that is being written. This allows other plug-ins to transmit images for visualization or for other plug-ins but not write them to file. For example, this is useful in the standard plane detection plug-in, where the user may not want to write the `background' images to file, but still wants to visualise them in real time.

This plug-in implements a widget that shows the number of images that have been saved and allows to stop/resume the image saving via a checkbox.

\subsection{GUI plug-in}
\label{sec:plugin:gui}

\begin{figure}[htb!]
    \centering
    \includegraphics[width=0.8\linewidth]{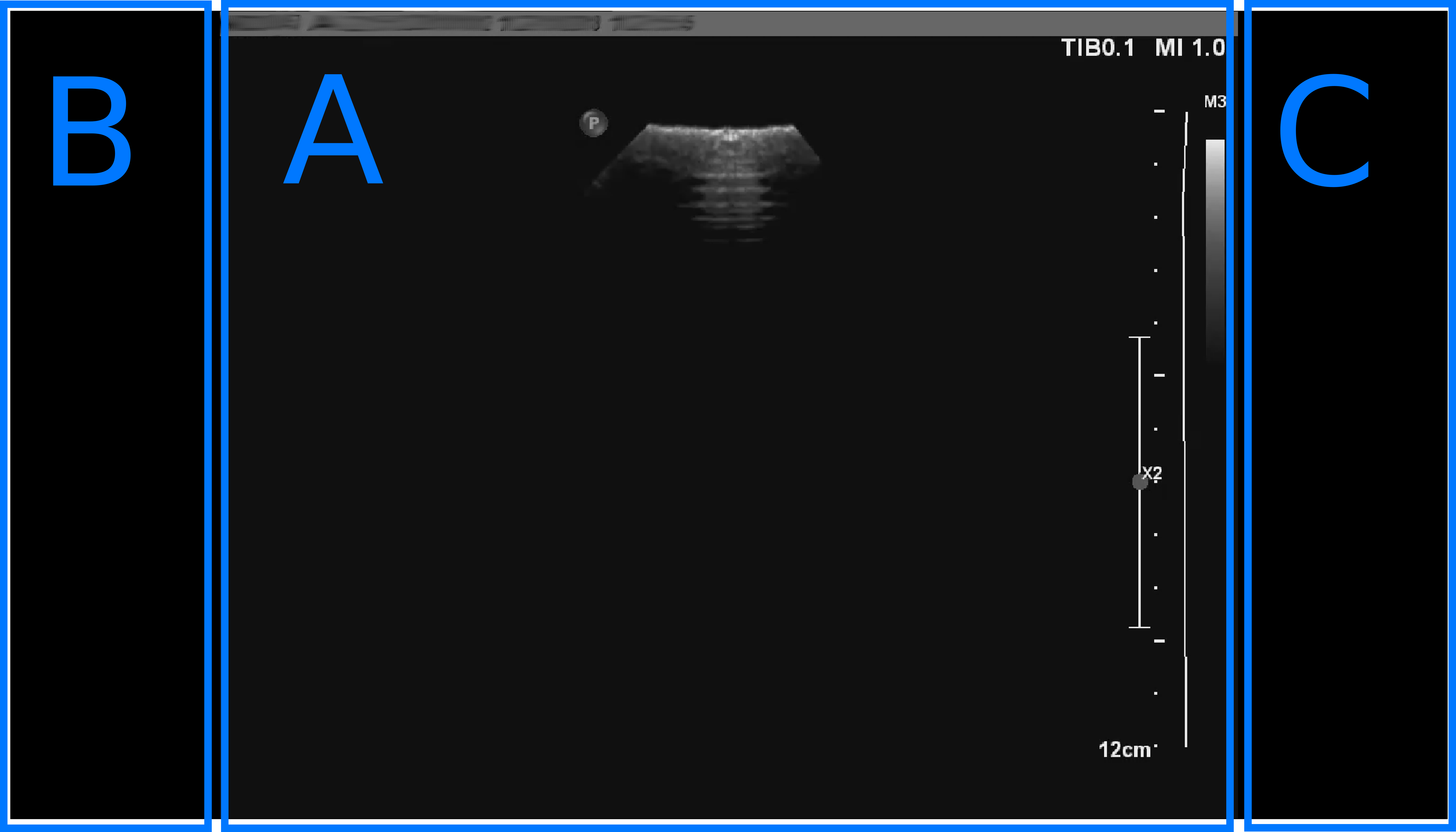}
    \caption{Interface of the visualization plug-in. The central frame (A) displays the images from a given stream in real time. The side frames (B and C) can be used to display widgets from individual plugins upstream the pipeline. }
    \label{fig:visualization-plugin}
\end{figure}

The graphical user interface (GUI) plug-in is designed to display a stream of images and widgets around the images with information of the other plug-ins in the pipeline. The organisation of the visualization window in shown in Figure~\ref{fig:visualization-plugin}.

All plug-ins can implement two types of widgets, declared in the \texttt{Plugin} parent class: plug-in widgets, that can be placed in panels B or C in the figure, and image widgets, that can be placed in panel A. By default, the GUI plug-in creates a coloured frame around each widget that matches a coloured frame around the image widget of the same plug-in (if available), as shown in Figure~\ref{fig:plugins}. This can be disabled with the command line argument \texttt{--gui\_usecolors 0}.

The GUI plug-in itself implements a widget (by default located in panel B) that allows to control the size of all image widgets.




\section*{Current executable software version}


\begin{table}[!h]
\begin{tabular}{|l|p{6.5cm}|p{6.5cm}|}
\hline
\textbf{Nr.} & \textbf{(Executable) software metadata description} & \textbf{Please fill in this column} \\
\hline
S1 & Current software version & 1.0 \\
\hline
S2 & Permanent link to executables of this version  &  $https://github.com/gomezalberto/$ $pretus/releases/tag/v1.0$  \\
\hline
S3 & Legal Software License & MIT license \\
\hline
S4 & Computing platforms/Operating Systems & Linux \\
\hline
S5 & Installation requirements \& dependencies & Qt $\geq$ 5.12, VTK $\geq$ 8.0, ITK $\geq$ 4.12, Boost. For Python plug-ins, in addition: PyBind11, Pyhton $\geq$ 3.6, numpy. Different plug-ins might have added dependencies, please check each plugin's repository.  \\
\hline
S6 & If available, link to user manual - if formally published include a reference to the publication in the reference list & 
\\
\hline
S7 & Support email for questions & pretus@googlegroups.com\\
\hline
\end{tabular}
\end{table}

\end{document}